# Dipole-Magnetic Photo-Recombination in Ultracold Hydrogen Plasma

(Short title: Dipole-Magnetic Photo-Recombination)

Arkadiy S. Baltenkov
Institute of Ion-Plasma and Laser Technologies, Uzbek Academy of Sciences
100125, Tashkent, Uzbekistan

**Abstract.** The cross section of dipole-magnetic photodisintegration of the negative hydrogen ion has been calculated within the framework of an effective-range theory. The dipole-magnetic cross section of photodetachment within the very narrow range of energy near the process threshold dominates over the dipole-electric one. It has been shown that in ultracold hydrogen plasma at temperatures below $kT=3.29 \cdot 10^{-4}$ K the dipole-magnetic photo-recombination becomes a main mechanism of electron capture by hydrogen atoms.

1. Photoelectron detachment of negative hydrogen ion was theoretically studied in many papers (see, for example, [1-3] and references therein). In those papers the photodetachment process was considered as electric-dipole absorption of photons. Near the process threshold the electric-dipole photodetachment of $s$-states of ion make no contribution to the cross section. According to the Wigner law [4], in the systems bounded by the short-range forces near the threshold the effect should be made by the electron transitions into the continuum $s$-state, i.e. the transitions that there are no in electric-dipole absorption. They are absent also in the electric-quadrupole photo-process. Therefore, for the cross section of negative ion photodetachment (or that of the inverse one) to be calculated near the threshold, it is necessary to consider magnetic-dipole interaction allowing the transitions between the $s$-states of the discrete and continuum spectra.

   A necessity of considering these processes was demonstrated, for the first time, by Fermi in [5] where the recombination of neutron and proton with deuteron formation and photon emission was studied. This process is that inverse to deuteron photodetachment and the cross sections of these processes are connected with each other by the detailed equilibrium principle. Near the threshold for small momentums $p$ of relative motion of proton and neutron the dipole-electric cross section of radiation capture is in direct proportional to $p$ while the dipole-magnetic cross section is inversely to this momentum. For this reason, the probability of slow neutron capture by proton is independent of speed of their relative motion in the magnetic process and decreases as $p^2$ in the dipole-electric process. In the diagrams of the deuteron photo-electric and photo-magnetic detachments the latter process manifests itself as a small peak at the deuteron disintegration [6].

   Bearing all of this in mind, some questions arise. What is the role of the magnetic field of the light wave in the processes of negative ion photodetachment, in particular, hydrogen ions? What is the range of photoelectron energy where the magnetic mechanism of photodetachment (or photo-recombination) dominates over the electric one? This paper is devoted to these questions.

2. In the dipole approximation the cross section of photon absorption by extra electron of the ion H⁻ is given by the formula [7, 8]

$$\sigma(\omega) = \frac{4\pi}{3} \frac{e^2}{\hbar c} \frac{\omega m k}{2\pi \hbar} |M|^2, \qquad (1)$$

where $m$ and $e$ are the electron mass and charge; $k$ is the wave vector of knocked-out electron; $c$ is the speed of light; $\hbar\omega$ is the photon energy; $M$ is the matrix element of dipole-electric or dipole-magnetic moments. These matrix elements are the following integrals

$$M_{el} = \langle \mathbf{k}\uparrow|z|i\uparrow\rangle = \int \psi_\mathbf{k}^{\uparrow*}(\mathbf{r}) z \psi_i^{\uparrow}(\mathbf{r}) d\mathbf{r}, \qquad (2)$$

$$M_{mag} = \frac{\lambda}{2}\langle \mathbf{k}\downarrow|i\uparrow\rangle = \frac{\lambda}{2}\int \psi_\mathbf{k}^{\downarrow*}(\mathbf{r}) \psi_i^{\uparrow}(\mathbf{r}) d\mathbf{r}. \qquad (3)$$

Here $\lambda = \hbar/mc$ is the Compton length of the electron wave; $e\lambda/2$ is the magnetic moment of electron. The arrows at the wave functions indicate the directions of extra electron spin in the bound $^1S_0$ state and in the continuum. The electric interaction (2) leads to the electron dipole transitions from the $s$-ground state to the $p$-continuum; the H⁻ ion remaining in the $^1S_0$ state. The magnetic interaction (3) leads to the electron transitions from the ion ground $s$-state to the $s$-state in continuum. This transition is accompanied by electron spin-flip. The ion goes from the singlet $^1S_0$ state to the triplet $^3S_1$ one. Since the $s$-state is the final one, the photoelectrons have the isotropic distribution in angles in contrary to photoelectric detachment for which the $p$-state is final. Because of the absence of interference between the photoelectron partial waves generated in the electric and magnetic processes the total cross section of negative ion photodetachment is a sum of the cross sections of dipole-electric and dipole-magnetic photodetachment

$$\sigma_{tot}(\omega) = \sigma_{el}(\omega) + \sigma_{mag}(\omega) \qquad (4)$$

Bethe and Longmire [9] have discussed the photodisintegration of the deuteron essentially based on the loosely bound nature of the deuteron. Their arguments can be applied to the bound-free extra electron transition from the negative hydrogen ion in which the radius of the electron orbit significantly exceeds the radius of the parent H atom. Within the effective range theory [9] the wave functions of extra electron of ion in the integrals (2) and (3) can be written in the following forms

$$\psi_i^{\uparrow}(\mathbf{r}) = \sqrt{\frac{\kappa}{2\pi(1-\kappa\rho_s)}} \frac{e^{-\kappa r}}{r}, \qquad (5)$$

$$\psi_\mathbf{k}^{\uparrow}(\mathbf{r}) = e^{i\mathbf{k}\cdot\mathbf{r}} - \frac{\sin kr}{kr} + \frac{\sin(kr+\delta_0^{\uparrow})}{kr}, \qquad (6)$$

$$\psi_\mathbf{k}^{\downarrow}(\mathbf{r}) = e^{i\mathbf{k}\cdot\mathbf{r}} - \frac{\sin kr}{kr} + \frac{\sin(kr+\delta_0^{\downarrow})}{kr}. \qquad (7)$$

Here $\rho_s$ is the effective radius of the ion in the singlet $^1S_0$ state; $\hbar^2\kappa^2/2m = I_s$ is the electron affinity of H⁻. The phases of the electron $s$-scattering in the singlet system $\delta_0^\uparrow \equiv \delta_0^s$ are described by the electron affinity of H⁻ and by the effective range $\rho_s$ of the ion state:

$$k \cot \delta_0^s = -\kappa + \frac{1}{2}\rho_s(\kappa^2 + k^2). \qquad (8)$$

The phases of the electron s-scattering in the triplet system $\delta_0^\downarrow \equiv \delta_0^t$ are connected with the scattering lengths and the effective ranges by the following formula

$$k \cot \delta_0^t = -\frac{1}{a_t(k)} = -\frac{1}{a_t} + \frac{1}{2}\rho_t k^2. \qquad (9)$$

The wave functions (5) and (6) with the same spins are orthogonal and the functions (5) and (7) with differently directed spins are, of course, nonorthogonal and so the overlapping integral (3) of these functions is nonzero.

The calculation result of the electric cross section is well known [2, 3]

$$\sigma_{el}(\omega) = \frac{16\pi}{3}\frac{e^2}{\hbar c}\frac{\hbar^2}{m}\frac{I_s^{1/2}(\hbar\omega - I_s)^{3/2}}{(\hbar\omega)^3}\frac{1}{(1-\kappa\rho_s)} \qquad (10)$$

For the magnetic cross section by means of simple manipulations with integral (3) we obtain the following formula

$$\sigma_{mag}(\omega) = \frac{2\pi}{3}\frac{e^2}{\hbar c}\lambda^2 \frac{[\kappa - 1/a_t(k)]^2 k\kappa}{[k^2 + 1/a_t^2(k)](k^2 + \kappa^2)}\frac{1}{(1-\kappa\rho_s)} \qquad (11)$$

If one introduces into consideration the energy of the virtual triplet state of the H⁻ ion $I_t = \hbar^2/2ma_t^2$ and omits the k-dependent terms in the formula (9), the magnetic cross section can be written in the same form as the electric one

$$\sigma_{mag}(\omega) = \frac{2\pi}{3}\frac{e^2}{\hbar c}\left(\frac{\hbar}{mc}\right)^2 \frac{I_s^{1/2}(I_s^{1/2} - I_t^{1/2})^2(\hbar\omega - I_s)^{1/2}}{(\hbar\omega + I_t - I_s)\hbar\omega}\frac{1}{(1-\kappa\rho_s)}. \qquad (12)$$

The both cross sections of photo-electric and photo-magnetic detachment decrease for high photon energy $\hbar\omega \gg I_s$ as $(\hbar\omega)^{-3/2}$. Near the threshold $\hbar\omega \to I_s$ the photo-electric cross section is proportional to $(\hbar\omega - I_s)^{3/2}$ while the photo-magnetic one to $(\hbar\omega - I_s)^{1/2}$ according to the general behavior of the cross sections near the reaction threshold [8].

3. The calculation results of the electric (10) and magnetic (12) cross sections are presented in Fig. 1. In these calculations it is assumed that $I_s = 0.754$ eV, $\rho_s = 2.646$, $I_t = 2.505$ eV [2]. The latter value corresponds to the triplet length of scattering $a_t = 2.33$ atomic units [2]. For the both cross sections to be presented in one figure, the magnetic cross section was magnified by 500 times. The electric cross section at maximum ($\hbar\omega = 2I_s$) has the value $\sigma_{el}^{max} \approx 41 \cdot 10^{-18}$ cm$^{-2}$ while the maximal value of the magnetic cross section is $\sigma_{mag}^{max} \approx 5.1 \cdot 10^{-22}$ cm$^{-2}$. The smallness of the ratio $\sigma_{mag}^{max}/\sigma_{el}^{max} \approx 10^{-5}$ is due to the fact that the magnetic dipole moment $e\lambda/2$ is small as compared to the electric one $ea$ because the radius of extra electron orbit $a$ is much more than the Compton length of the electron wave $\lambda$. For photon energy very close to the photodetachment threshold, as seen in Fig. 1, the photo-magnetic cross section exceeds the electric one. This energy range is very narrow to fix the magnetic photodetachment. However, under some conditions this effect becomes determining in the inverse process.

4. The radiation capture of free electron by the hydrogen atom with emission of recombination photon is the process inverse to negative ion photodetachment. The recombination cross sections are connected with those of photodetachment (10) and (12) by the principle of detailed equilibrium [8] according to which

$$\sigma_{el}^{rec}(\omega) = 2\sigma_{el}(\omega)\left(\frac{\omega}{ck}\right)^2; \quad \sigma_{mag}^{rec}(\omega) = 2\sigma_{mag}(\omega)\left(\frac{\omega}{ck}\right)^2 \qquad (13)$$

Here $\hbar\omega$ is the recombination photon energy; $k$ is the wave vector of captured electron. The ratio of these cross sections near the threshold is

$$\frac{\sigma_{mag}^{rec}}{\sigma_{el}^{rec}} = \frac{\alpha^2}{8}\frac{(I_s^{1/2}-I_t^{1/2})^2}{I_t}\frac{I_s}{2Ry}\frac{I_s}{(\omega-I_s)}. \qquad (14)$$

Here $\alpha = e^2/\hbar c$ is the fine structure constant; $2Ry$ is the atomic unit of energy. Setting this ratio equal to unity, we find that for electron energies less than $\varepsilon = (\hbar\omega - I_s) = mv^2/2 = 2.836\cdot 10^{-8}$ eV the magnetic cross section exceeds the electric one. Thus, for plasma temperature lower than kT=3.29·10$^{-4}$ K the magnetic process becomes a main mechanism of radiation recombination of free electrons with hydrogen atoms in partially ionized ultracold plasma where the three-body recombination, which is prevalent in usual plasmas, strongly suppressed [10].

The cross section of magnetic recombination at the threshold is inversely proportional to the electron speed

$$\sigma_{mag}^{rec}(\omega \approx I_s) = \frac{2\pi}{3}\alpha\lambda^4\frac{(I_s^{1/2}-I_t^{1/2})^2 I_s}{I_t}\frac{1}{(1-\kappa\rho_s)}\frac{\kappa}{\hbar v}, \qquad (15)$$

while the electric cross section is directly proportional to the electron speed so it vanishes at the threshold. From (15) it follows that in ultracold plasma the free electron lifetime $\tau$ due to the magnetic process is independent of temperature and defined by the following constant

$$n_a\tau = \frac{1}{\sigma_{mag}^{rec}v} = \frac{3I_t(1-\kappa\rho_s)}{2\pi\alpha(I_s^{1/2}-I_t^{1/2})^2}\frac{\hbar}{\lambda^4\kappa I_s} \approx 1.07\cdot 10^{21} \text{ sec/cm}^3. \qquad (16)$$

Here $n_a$ is the concentration of neutral hydrogen atoms.

Considered here, the mechanism of radiation recombination perhaps will allow explaining the behavior of free electrons in ultracold plasmas, which exist in earthly labs [11, 12] and do occur naturally in astrophysical systems like white dwarf stars or interior of gas giant planets.

**Acknowledgments**
The author is grateful to Profs. S. Manson and A. Msezane for useful discussion. This work was supported by the Uzbek Foundation Award Ф2-ФА-Ф164.

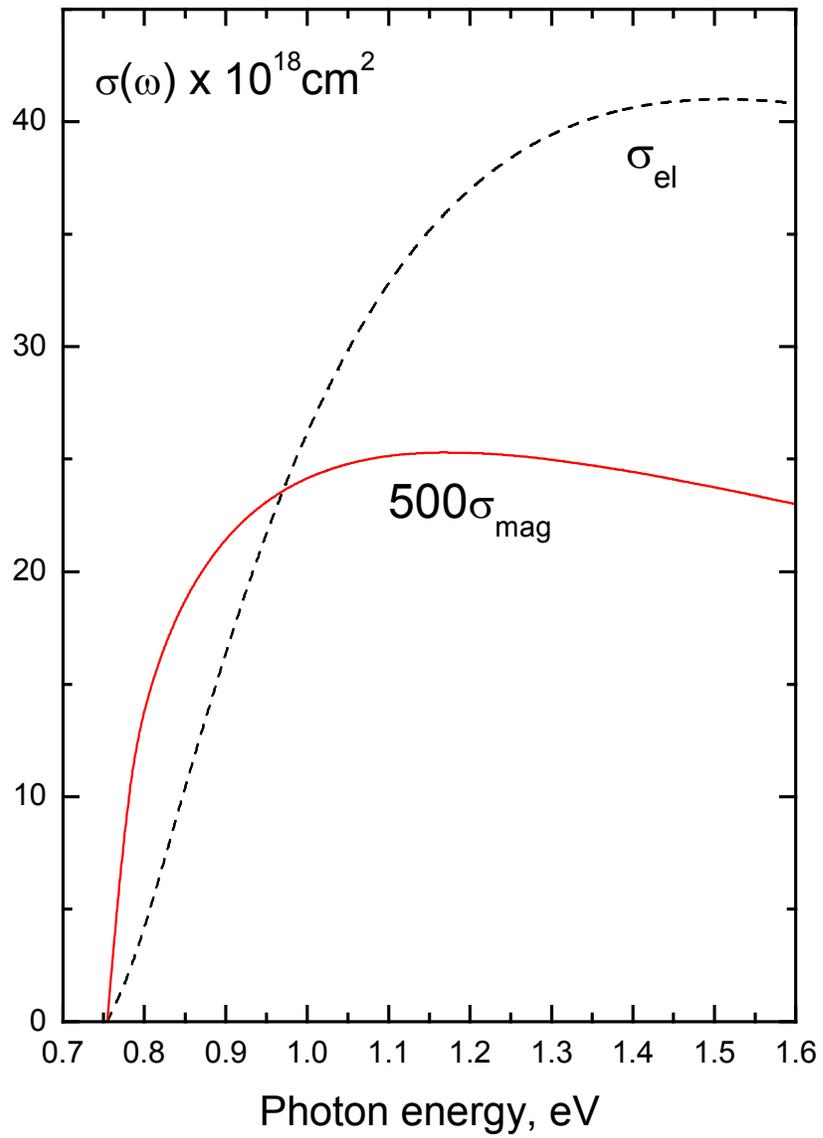

Fig. 1. Photodetachment cross sections.